# Asymmetry in the Spectrum of High-Velocity H$_2$O Maser Emission Features in Active Galactic Nuclei

A. V. Nesterenok[*] and D. A. Varshalovich[**]

*Ioffe Physical - Technical Institute, ul. Politekhnicheskaya 26, St. Petersburg, 194021 Russia*

Abstract **-** We suggest a mechanism for the amplification of high-velocity water-vapor maser emission features from the central regions of active galactic nuclei. The model of an emitting accretion disk is considered. The high-velocity emission features originate in the right and left wings of the Keplerian disk. The hyperfine splitting of the signal levels leads to an asymmetry in the spectral profile of the water vapor maser line at a frequency of 22.235 GHz. We show that the gain profile asymmetry must lead to an enhanced brightness of the blueshifted high-velocity emission features compared to the redshifted ones. Such a situation is observed in the source UGC 3789.

## INTRODUCTION

Investigation of the maser emission that originates in accretion disks in the central regions of galaxies opens a unique opportunity to directly measure the Hubble constant. The high intensity of the maser emission allows the spatial structure of the system of sources to be investigated by radio-interferometry methods. The narrowness of the spectral features makes it possible to measure the frequencies of emission lines with a high accuracy. This allows one to model the geometry and kinematics of the accretion disk and to determine the "geometric" distance to the system. The accuracy of determining cosmological parameters, such as the Hubble constant, depends significantly on the calibration of the intergalactic distance scale. At present, this scale is based on observations of Cepheids in the Large Magellanic Cloud and in the nearest galaxies. Since the distance to the Large Magellanic Cloud is uncertain (its independent estimates can differ by as much as 25%) and since the effect of metallicity on the period – luminosity relation for Cepheids is unclear, the accuracy of distance determination by this method is limited (Macri et al. 2006). One way of refining the intergalactic distance scale is to use the systems whose distances can be found by "geometric" methods as "reference points". One of such systems is the galaxy NGC 4258, whose geometric distance was obtained on the basis of detailed interferometric measurements and by modeling the geometry and kinematics of the accretion disk around a central black hole.

---

[*] E-mail: alex-n10@yandex.ru
[**] E-mail: varsh@astro.ioffe.ru

The $H_2O$ maser emission from the central region of NGC 4258 was detected by Claussen et al. (1984); the emission frequency shift corresponds to the galaxy's recession velocity of about 500 km s$^{-1}$, and the "isotropic luminosity" in the 1.35-cm maser line (22.235 GHz) is 120 $L_\odot$. Apart from the brightest central source, Nakai et al. (1993) detected less intense, high-velocity emission features with velocities relative to the central source of about ±1000 km s$^{-1}$. Miyoshi et al. (1995) were the first to map the spatial distribution of maser sources by means of VLBI. As a result, it was established that the maser emission from the central region of NGC 4258 originates in a thin, differentially rotating accretion disk seen edge-on. The emission whose frequency corresponds to the systemic velocity of the galaxy comes to us from the disk foreground, where the rotation velocity of the emitting gas is perpendicular to the line of sight. The high-velocity maser emission features originate in the right and left disk sectors, where the orbital velocity of the emitting gas is parallel and antiparallel to the line of sight. The orbital radii of the high-velocity features lie within the range from 0.11 to 0.3 pc, and their velocities fall on a Keplerian rotation curve with an error of <1% (Argon et al. 2007). The central mass of the system calculated from the rotation curve is about $3.8 \times 10^7$ $M_\odot$ (Herrnstein et al. 2005).

The relative velocities of the spectral emission features in the central source undergo an acceleration of about 8 km s$^{-1}$ yr$^{-1}$, which corresponds to the centripetal acceleration of emitting gas clouds. Based on the interferogram data and the accelerations of the emission features in the central source, Herrnstein et al. (1999) determined the geometric distance to the system, 7.2 ± 0.3(rand.) ± 0.4(syst.) Mpc. A further refinement of the geometry of the spatial distribution of maser sources and their accelerations will allow the geometric distance to the system to be determined with an accuracy as high as 3% (Argon et al. 2007; Humphreys et al. 2008).

Since the distance to NGC 4258 is relatively small (7.2 Mpc), its peculiar velocity is significant compared to the cosmological recession velocity of the galaxy. Furthermore, the gravitational frequency shift due to the difference in the gravitational potentials of the central region of NGC 4258 and our Galaxy can also be significant. All of these factors hampered the direct determination of the Hubble constant, given the derived geometric distance to the galaxy. Nevertheless, this would be possible for a system like NGC 4258 but located at greater cosmological distances.

Braatz and Gugliucci (2008) reported the discovery of water-vapor maser emission in the central region of the galaxy UGC 3789 with a cosmological velocity of 3325 km s$^{-1}$. At a Hubble constant of 72 km s$^{-1}$ Mpc$^{-1}$, the corresponding distance to UGC 3789 is about 46 Mpc. The peculiarities of the maser emission spectrum for UGC 3789 suggest that, as in the case of NGC 4258, the emission is generated in a differentially rotating accretion disk (Reid et al. 2008). Apart from the main maser source at a velocity close to the systemic velocity of the galaxy, high-velocity subsources are observed whose relative velocities as a function of the distance to the system's center fall on a Keplerian rotation curve. The rotation radii of the blueshifted high-velocity sources lie within the range from 0.08 to 0.16 pc, and the rotation velocities reach 800 km s$^{-1}$. The redshifted sources extend to distances from 0.11 to 0.3 pc from the center and have velocities as high as 650 km s$^{-1}$. The mass of the system's central body determined from the rotation curve is about $1.1 \times 10^7 \, M_\odot$.

One of the peculiarities of the maser emission from NGC 4258 and UGC 3789 is an asymmetry in the spectrum of the high-velocity emission features. In the system NGC 4258, the energy flux density of the emission from the redshifted high-velocity sources is more than an order of magnitude higher than that from the blueshifted ones. Based on the accretion disk model of Neufeld and Maloney (1995), Herrnstein et al. (1996) suggested that the asymmetry in the emission spectrum was due to the orientation of the disk warps with respect to the line of sight. In their model, the emission of the blueshifted high-velocity subsources passes through a layer of ionized gas and undergoes partial absorption on free electrons. In contrast, for UGC 3789, the blueshifted maser subsources are most intense in the system of masers. The asymmetry in the spectrum of the high-velocity emission features reflects the fact that the physical conditions in the left and right disk wings are not identical. An understanding of these conditions is needed to construct a more accurate accretion disk model.

In this paper, we present the results of our calculation of the intensities for the side high-velocity maser emission features by taking into account the hyperfine splitting of the signal levels. We show that for small inclinations of the disk plane to the line of sight, a gain profile asymmetry must lead to an enhanced brightness of the blueshifted high-velocity maser features.

## SPECTRAL LINE PROFILE WITH THE HYPERFINE SPLITTING

Since the nuclear spin of oxygen $^{16}$O is zero and the nuclear spin of the proton is 1/2, there exist two states of the water molecule, para- and ortho- states, with a total nuclear spin $I$ of zero and one, respectively. The levels of para-H$_2$O molecules have no hyperfine splitting, while the levels of ortho-H$_2$O molecules are split into three sublevels, $F = J - 1$, $J$, and $J + 1$, where $J$ is the rotational angular momentum of the molecule and $\boldsymbol{F} = \boldsymbol{J} + \boldsymbol{I}$ is the total angular momentum of the system.

The maser line at $\lambda = 1.35$ cm ($\nu = 22.235$ GHz) corresponds to the permitted electric dipole transition between the rotational levels of ortho-$H_2O$ molecules $J_{KaKc} = 6_{16} \rightarrow 5_{23}$ ($K_a$ and $K_c$ are the asymptotic quantum numbers that characterize the components of vector $\mathbf{J}$ along the inner axes of the molecule). For the transition under consideration, the spectral gain profile is the sum of six components with different intensities. Only three components are the most intense ones. According to Varshalovich et al. (2006), the expression for the spectral profile of the gain $\varphi_{HF}(u)$ is

$$\varphi_{HF}(u) = ae^{-\left(\frac{u-u_a}{v_T}\right)^2} + be^{-\left(\frac{u-u_b}{v_T}\right)^2} + ce^{-\left(\frac{u-u_c}{v_T}\right)^2},$$

where $u$ is the relative velocity related to the emission frequency $\nu$ by $u = -c\left(\frac{\nu-\nu_0}{\nu_0}\right)$, $\nu_0 = $ 22.235044 GHz is the frequency of the most intense splitting component, the quantity $v_T$ characterizes the thermal spread in velocities of the emitting molecules, $v_T = \sqrt{\frac{2kT}{m}}$, and the coefficients have the following values: $a = 0.3919$; $b = 0.3302$; $c = 0.2779$; and $u_a = 0$; $u_b = -0.4458$; $u_c = -1.0293$ km s$^{-1}$. The profile is normalized in such a way that $\int \varphi_{HF}(\nu) d\nu = \sqrt{\pi}\Delta\nu_D$ where $\Delta\nu_D = \nu_{J''J'} v_T/c$, and $\nu_{J''J'}$ - is the mean transition frequency. Figure 1 presents the spectral profile of $\varphi_{HF}(u)$ for temperatures $T = 50, 150, 300, 600$, and 1000 K. The FWHM of the gain profile, given its multicomponent structure, varies from 1 to 2 km s$^{-1}$. However, the observed maser lines can be considerably narrower (see, e.g., Lekht and Richards 2003).

## RELATIVISTIC EFFECTS

The relativistic corrections to the classical Doppler effect for the redshifted and blueshifted high-velocity maser features are $V^2/2c = 1.7$ km s$^{-1}$ for a rotation velocity of 1000 km s$^{-1}$ (see Eq. (3) below). The frequency shift of the emission from the central source of NGC 4258 due to the transverse Doppler effect is the same in absolute value but opposite in sign. The corresponding shifts for the system UGC 3789 do not exceed 1 km s$^{-1}$ in absolute value.

The gravitational redshift of maser lines due to the gravitational potential of the central body for the system NGC 4258 is $GM/(Rc) = 5$ km s$^{-1}$ for an orbit with a radius of 0.11 pc and 2 km s$^{-1}$ for an orbit with a radius of 0.29 pc; for the system UGC 3789, it is 1.6 km s$^{-1}$ for an orbit with a radius of 0.1 pc.

The relativistic corrections to the velocities of Keplerian orbits around a rotating black hole $\Delta v/v$ do not exceed $GM/Rc^2 \approx 10^{-5}$ (Bardin et al. 1972) and, hence, are insignificant.

According to our estimates in the problem under consideration, the refraction of the emission in a differentially rotating disk with a nonuniform field of the refractive index is negligible.

CALCULATION OF THE INTENSITIES OF HIGH-VELOCITY EMISSION FEATURES.

Let us consider the model of an accretion disk whose plane is parallel to the line of sight. The emission intensity in a homogeneous medium is

$$I_v(\tau) = I_v(0) \cdot e^{-\tau} + S_v(1 - e^{-\tau}),$$

where $S_v$ is the source function and $\tau$ is an optical depth at frequency $v$. In the amplification regime $\tau < 0$, $S_v < 0$, and $I_v(\tau) \propto e^{|\tau|}$. The optical depth $\tau$ at frequency $v$ is an integral of the gain $\kappa(v)$ along the line of sight:

$$|\tau| = \int \kappa(v) dl.$$

In the case of isotropic pumping of the signal levels and in the rest frame of reference, the expression for the gain $\kappa(v)$ is (Varshalovich et at. 2006):

$$\kappa(v) = \frac{4}{3}\pi^{3/2} \frac{S(J'' \to J')}{\hbar} \cdot \frac{N}{v_T} \cdot \left( \frac{n_J''}{2J''+1} - \frac{n_J'}{2J'+1} \right) \cdot \varphi_{HF}(v), \quad (1)$$

where $J''$ and $J'$ are the quantum numbers of angular momentum for the upper and lower maser levels; $S(J'' \to J')$ is the line strength of the transition under consideration; $N$ is the number density of working molecules; $n_J''$ and $n_J'$ are the populations of the upper and lower levels, respectively, normalized according to $\sum_J n_J = 1$, and $\varphi_{HF}(v)$ is the spectral profile of the gain with the hyperfine splitting.

The emitted line frequencies undergo a shift due to general relativity effects, namely, a cosmological redshift, a shift due to the difference between the gravitational potentials of the galaxy under study and our Galaxy, the gravitational potential of the central massive body, and a shift due to the kinematic Doppler effect. For an observer in the galaxy under study, the emission frequency is defined by the expression

$$v_0' = v_0 \sqrt{1 - \frac{2GM}{Rc^2}} \frac{\sqrt{1 - V^2/c^2}}{1 + \frac{V}{c}\cos\theta}, \quad (2)$$

where $v_0$ is the emitted frequency, $M$ is the mass of the central body, $R$ is the distance from the system's center to the source, $V$ is the source's velocity, and $\theta$ is the angle between the velocity vector $V$ and the line of sight; for the redshifted and blueshifted high-velocity sources, $\theta \approx 0$ and $\theta \approx \pi$, respectively. Since the gas moves in circular Keplerian orbits in the model under consideration, $\frac{GM}{Rc^2} = \frac{V^2}{c^2}$, i.e., the gravitational redshift has the second order of smallness in $V/c$. Expanding (2) in a series of $V/c$ and retaining the terms up to $V^2/c^2$, we have

$$v'_0 = v_0 \left(1 - \frac{V}{c}\cos\theta + \frac{V^2}{c^2}\left(\cos^2\theta - \frac{1}{2}\right) - \frac{V^2}{c^2}\right), \qquad (3)$$

where the last term in brackets corresponds to the gravitational redshift. Including the mentioned effects in Eq. (1) for the gain is reduced to substituting $v'_0$ for $v_0$. When we pass to a moving frame of reference, the spectral line profile is shifted in frequency, virtually without changing its shape. The emergence of the emission from resonance and, hence, the coherent emission length are determined by the change in $\frac{V}{c}\cos\theta$ along the line of sight. The quadratic term has the same sign for the redshifted and blueshifted sources and determines their identical additional shift in frequency.

Figure 2 presents the calculated intensities of the redshifted and blueshifted high-velocity spectral maser features for temperatures of 150, 300, 600, and 1000 K. This temperature range is considered as one of the necessary conditions for the generation of water-vapor maser emission (see, e.g., Babkovskaya and Varshalovich 2000; Argon et al. 2007). The orbital velocity of 763.6 km s$^{-1}$ for which the calculations were performed is marked in the figure by the vertical line. This velocity corresponds to an orbital radius of 0.28 pc in the system NGC 4258 or to an orbital radius of 0.1 pc in the system UGC 3789. In our calculations, we disregarded the radial dependence of the parameters, $n_{J'}$, $n_{J''}$ and $N$, i.e., we assumed that the emission of an individual high-velocity maser source was amplified almost locally. The observed intensities of the high-velocity emission features correspond to $|\tau| \approx 25$.

DISCUSSION

According to the data presented in Fig. 2, the blueshifted high-velocity spectral emission features are amplified more efficiently than the redshifted ones. This effect is due to the different orientation of the spectral gain profile relative to the direction of the emission frequency shift as the gas moves along a circumference (see Fig. 3). The amplification efficiency of the emission from the blueshifted high-velocity sources relative to the redshifted ones decreases with increasing temperature: the profile becomes more symmetric as the temperature increases (see Fig. 1). The result also depends on the optical depth $\tau$ of the emission being amplified: the difference in the optical depths in the high-velocity emission features is proportional to the absolute value of $\tau$.

It follows from Fig. 2 that the relative shift in the absolute values of the velocities for the redshifted and blueshifted spectral emission features is 3 km s$^{-1}$: the relativistic correction to the classical Doppler effect for this rotation velocity is 1 km s$^{-1}$ for each component; the additional shift of 1 km s$^{-1}$ is due to an asymmetry in the gain profile relative to the "zero" frequency $v_0$ of the most intense splitting component (see Figs. 1 and 3). The gravitational redshift of lines, about 2 km s$^{-1}$, is not reflected in Fig. 2.

In the model under consideration, the coherent length of the high-velocity maser features $l$, the distance at which the emission emerges from resonance by shifting by $\Delta v_D$, is ~$10^{16}$ cm. Since the disk thickness is $h = \alpha r$, where $\alpha \approx v_T / V$ and $V$ is the orbital rotation velocity of the gas, we obtain a condition on $i$, the disk inclination with respect to the line of sight, at which the gain profile asymmetry affects the intensity of the features: $|i| < \alpha r/l$. For the parameters of the systems NGC 4258 and UGC 3789, we have $\alpha r/l \approx 0.02$ or 1°. At values of $i$ that are appreciably higher than this limiting value, there is no difference in amplification of the features and the intensity profile for an individual maser line reflects the asymmetry of the spectral gain profile.

It is worth noting that in an actual warped disk, the parameter $i$ can take on various values. For example, based on the space-velocity distribution of subsources in the system NGC 4258, Herrnstein et al. (2005) suggested an accretion disk model in which the inclination of the disk plane to the line of sight $i$ for the redshifted high-velocity maser features changed from −9° for a radius of 0.13 pc (the observer sees the upper plane of the disk) to +1° for a radius of 0.29 pc (the observer sees the lower plane of the disk). The situation is similar for the blueshifted high-velocity maser features. At distances of about 0.28 pc from the center, the disk inclination with respect to the line of sight is close to zero and the condition for relative amplification of the blueshifted high-velocity emission features is met. However, according to experimental data, the reverse is true: in the system NGC 4258 the redshifted spectral emission features are more intense. Based on the spatial distribution of maser points in the central source, Argon et al. (2007) obtained an upper limit for the temperature of the disk material, 600 K. For such temperatures, the amplification of the blueshifted high-velocity features relative to the redshifted ones in the model under consideration is inefficient (see Fig. 2). The observed relative intensity of the redshifted spectral emission features can be caused by the mechanism proposed by Herrnstein et al. (1996).

## CONCLUSIONS

The hyperfine splitting of the signal levels for the $H_2O$ maser line leads to an additional broadening of the spectral gain profile and to its asymmetry. We showed that the profile asymmetry could be one of the causes for the enhanced brightness of the blueshifted high-velocity emission features observed in the system UGC 3789 with respect to the redshifted ones. In addition, the profile asymmetry relative to the "zero" emission frequency leads to an additional line frequency shift comparable to the relativistic corrections to the classical Doppler effect in the problem under consideration.

## ACKNOWLEDGMENTS

This work was supported by the Russian Foundation for Basic Research (project no. 08-02-01246-a), the Program of the President of Russia for Support of Scientific Schools (project no. NSH-3769.2010.2), and the "Active Processes in the Universe" Program of the Division of Physical Sciences of the Russian Academy of Sciences.

*Translated by G. Rudnitskii*

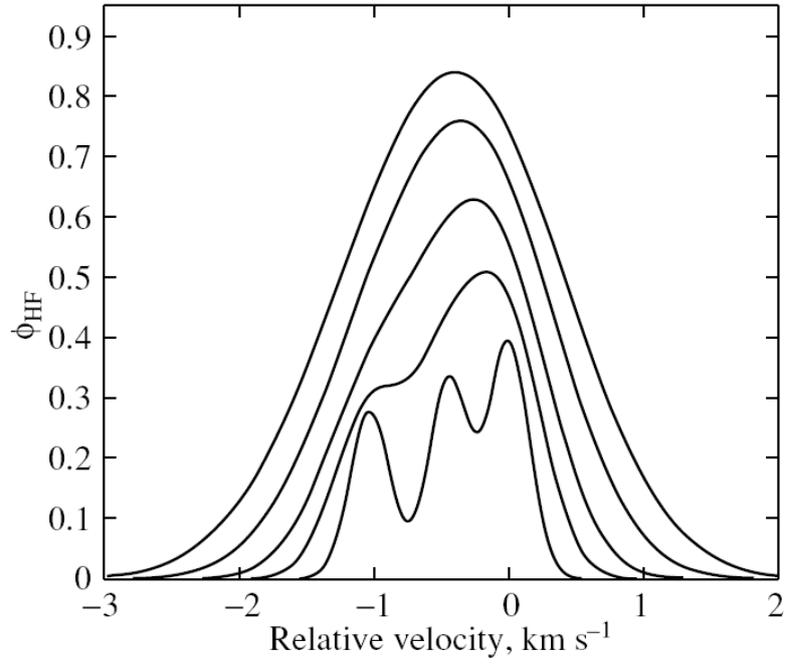

Fig. 1. Spectral gain profile for the water-vapor maser $6_{16} \to 5_{23}$ line with the hyperfine splitting of the signal levels for temperatures (from the bottom upward) of 50, 150, 300, 600, and 1000 K.

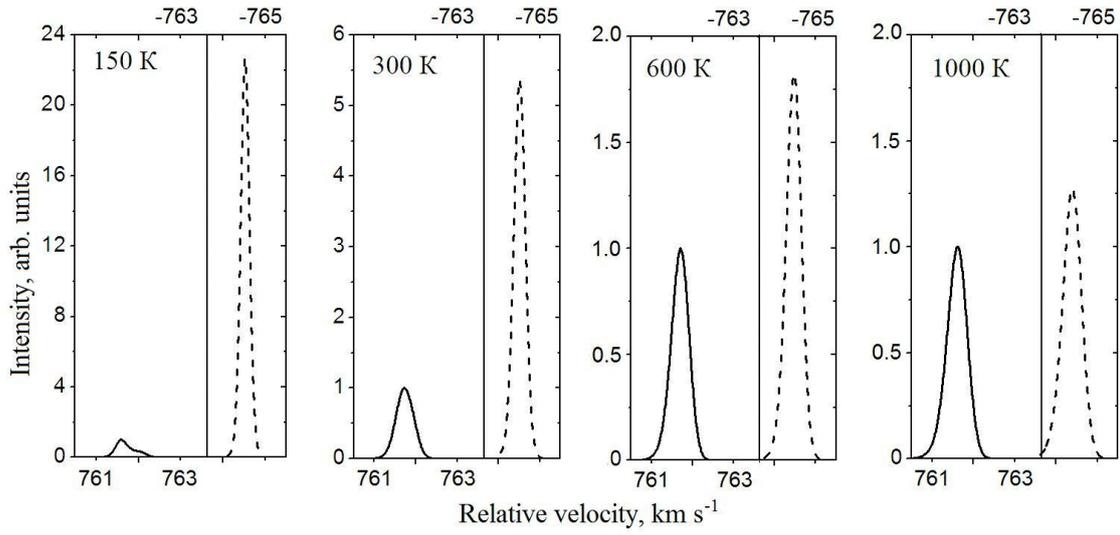

Fig. 2. Calculated intensities of the redshifted (solid line) and blueshifted (dashed line) high-velocity emission features for the model of a flat Keplerian disk. The relative velocity in the disk center-of-mass frame of reference is along the horizontal axis; the lower and upper scales are for the redshifted and blueshifted sources, respectively. The vertical line marks the orbital velocity of 763.6 km s$^{-1}$ used in our calculations. The intensity in arbitrary units is along the vertical axis; the intensity of the redshifted emission feature was normalized to unity.

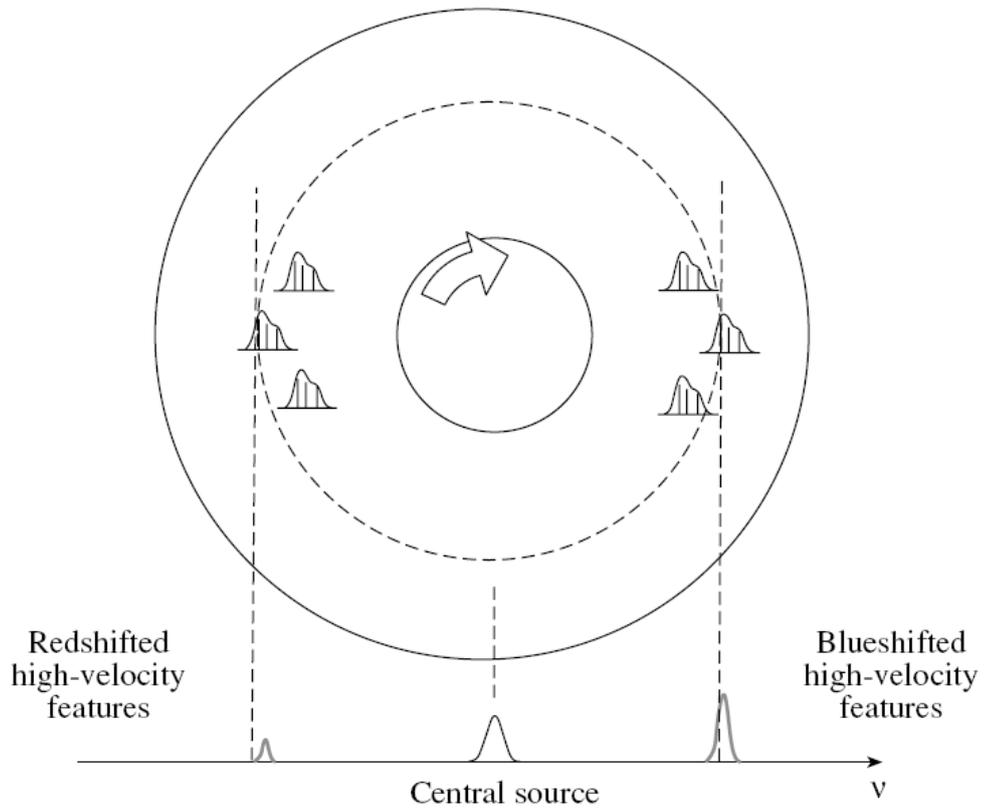

Fig. 3. Scheme demonstrating the change in emission frequency of the gas as it moves along a circumference; the direction of rotation is indicated by the arrow. An asymmetry in the spectral gain profile relative to the "zero" frequency of the most intense splitting component leads to an additional frequency shift of the high-velocity emission features. The figure is not to scale.